\documentclass[12pt]{article}
\usepackage{natbib}
\usepackage{graphics}
\usepackage[pdftex,dvips,xdvi]{graphicx}
\usepackage{setspace}
\usepackage{amssymb}\usepackage{rotate}
\usepackage{psfrag}
\usepackage{a4}

\title{Price Clustering and Discreteness: Is there Chaos behind the Noise?}

\author{Antonios Antoniou \and Constantinos E. Vorlow\thanks{Corresponding author email:
\texttt{Costas@vorlow.org}, www: \texttt{http://www.vorlow.org}. We
wish to thank Timothy Crack, James B. Ramsey, Tassos Malliaris,
Alexandros Leontitsis, and the participants of the APFA 2004
conference in Warsaw for their useful comments and suggestions on an
initial draft and earlier research which this paper advances. We
also wish to acknowledge the valuable help and support of Duncan
Rand and the University of Durham ``High Performance Computing
Service". We finally thank the anonymous referees of the 1st
Bonzenfreies Colloquium on Market Dynamics and Quantitative
Economics (2004), for their useful comments and suggestions. All
analysis was conducted on \texttt{R} (version 1.8.1, see
\cite{R:Ihaka+Gentleman:1996}; the software is available from
\texttt{http://www.r-project.org}).  The authors retain the sole
responsibility of any errors or omissions.}}

\date{\today}

\begin{document}

\maketitle

\begin{abstract}
We investigate the ``compass rose'' (Crack, T.F. and Ledoit, O.
(1996), Journal of Finance, 51(2), pg. 751-762) patterns revealed in
phase portraits (delay plots) of stock returns. The structures
observed in these diagrams have been attributed mainly to price
clustering and discreteness. Using wavelet based denoising, we
examine the noise-free versions of a set of FTSE100 stock returns
time series. We reveal evidence of non-periodic cyclical dynamics.
As a second stage we apply Surrogate Data Analysis on the original
and denoised stock returns. Our results suggest that there is a
strong nonlinear and possibly deterministic signature in the data
generating processes of the stock returns sequences.
\end{abstract}


\section{Introduction}

The empirical investigation of the dynamics of stock returns has
been an area of intensive research since the beginning of  last
century (see thesis of Bachelier \citep{Bachelier}). The
understanding of dynamics observed in price fluctuations are of
paramount importance to activities such as forecasting for
investment decision support, risk modelling and derivative pricing.
Moreover, the complexity of their structure, as a result of
agent-market interactions, is an indicator of the nature of overall
market conditions and organization. This complexity may also reflect
the level of agent's rationality and risk tolerance. It becomes
apparent that the explanation of certain qualities of the structure
of market dynamics, provides the opportunity to improve the
understanding of their current and future states. Clearly such an
exercise is of great importance to all market participants that aim
to minimize their risks and protect their investments and profits.

Viewing economies and markets in particular as a dynamical system,
we can draw many inferences by examining their observable outputs:
sequences of stock prices and the corresponding returns. Crack and
Ledoit \citep{Crack96} have first revealed  a ``\textit{compass
rose}" pattern discovered in scatter diagrams of returns against
their lagged values (i.e., phase portraits), such as the one
depicted in Fig. \ref{fig:fig1}(a). They attributed the pattern to
price clustering and discreteness and especially the tick size and
suggested reasons for its appearance.\footnote{Clustering in stock
market prices has been an issue that concerned research since the
1960's (e.g., see refs. \cite{Niederhoffer65,Niederhoffer66} who
were motivated by the original findings of \cite{Osborne62}). Ref.
\cite{Niederhoffer66b} investigated dependencies related to
clustering and discreteness. This research was followed by
\cite{Rosenfeld80} who conducted simulations on price rounding and
discreteness and showed that the hypothesis of a geometric Brownian
motion for daily and weekly frequencies could be rejected. In
general, price clustering and discreteness is an important chapter
of ``market's microstructure" with serious implications on risk
evaluation, the optimal design of securities and market efficiency.}
Our aim is by using an approach consistent with the tradition of
econophysics,  to continue their research by revealing yet more
interesting patterns and showing that the compass rose is a mask for
more subtle dynamics. In this paper we establish the case of
existence of nonstochastic nonlinear dynamics via the calculation of
the BDS statistic \citep{Brock87}. We use this as a discriminating
statistic for a permutation test based framework (``Surrogate Data
Analysis" (SDA) by \citep{Theiler92:surr}) that allows us to support
our results at various levels of significance. As a second step,
following \citep{ping:96,AntoniouVorlow:04,Wornel95}, we reduce the
level of noise in the original returns sequences using Wavelet based
thresholding (the Waveshrink technique by \citep{donoho95adapting}).
We then recalculate the BDS statistic on the denoised sequences and
their surrogates and test again for the absence of linear dynamics.
Meanwhile we produce the compass rose of the denoised sequences only
to reveal an entirely different structure that is strongly
reminiscent of a dynamical attractor. Our findings are consistent
with the hypothesis that the returns sequence dynamics may be
characterized by nonlinearities that can be of a
complex-deterministic character. The results produced here may bring
us closer to establishing that a significant part of the driving
force generating financial prices could indeed be chaotic.

\bigskip
\begin{center}
[ Insert figure \ref{fig:fig1} about here. ]
\end{center}
\bigskip

\section{Investigating the Compass Rose}\label{sec:data}

Crack and Ledoit \citep{Crack96} suggested first the use of phase
portraits in order to reveal the compass rose. This implied the
investigation of some sort of time-dependency among stock return
sequences. This could be linear or nonlinear, a result of stochastic
(random) or nonstochastic (deterministic) data generating process
(DGP), or even a mixture of the above behind the asset price
dynamics. The authors also proposed that the formations revealed
could be of use for calibrating tests of the existence of chaos in
returns sequences. Since then various papers have appeared on this
theme (see
\cite{Kramer97,Chen97,Szpiro98,Lee99,Gleason00,Wang00,Fang02,McKenzie:03}).
We believe that two issues can be addressed further:
\begin{enumerate}
    \item As \citep{Cho88} note, observed stock prices
    are not always the true equilibrium prices and hence the image of
    market dynamics observed through them could be partial. Moreover, in markets where
    significant fixing takes place, there is a variable amount of
    error introduced into the price level which is then passed to
    the returns (\citep[see][for a discussion on this]{Ball85}).
    \item Generating logarithmic or percentage returns, i.e., 1st
    order differencing, is a {\it high-pass}  filter \citep{abarbanel}. In this
    respect, all return sequences will contain amplified noise.
    Consequently, any interesting and possibly non-stochastic
    structures may be concealed and/or distorted. The importance
    of this becomes even greater if we take into account point (1)
    above.
\end{enumerate}

In the following pages, we investigate further the issue of compass
rose formations in stocks from the UK market. We analyze the daily
closing prices of stocks in the FTSE ALL SHARE and especially the
FTSE 100 index, spanning the period 01/01/1970 to 5/30/2003 (a
maximum of 8717 observations). A total of 53 FTSE100 stocks were
available with a full (homogeneous) range of prices for the above
time-span. Remarkably, all 53 high-capitalization company  prices
and corresponding returns revealed the patterns we observe and
report in this paper (some more intensively and clearly than
others).\footnote{ Although \citep{Crack96} use percentage returns,
we concentrate on continuously compounding return sequences
(logarithmic returns) and observe the same patterns.}

\section{Surrogate Data Analysis and Waveshrink}

Following \citep{Theiler92:surr,Schreiber00} (see also
\citep{Kaplan95,Kantz}), we investigated the possibility of the
observed structures of the compass rose being a ``one-off"
situation. The basic purpose of the SDA procedure is to provide a
framework that will allow us to deny the null hypothesis that the
data are generated by a linear stochastic system. It basically
comprises of two steps (see \citep{Galcka,Kantz} for an extensive
overview):
\begin{itemize}
\item The production of data sets from a model which captures deliberately only
certain ``linear" properties of the original sequence. These sets
are called ``surrogate data".
\item The rejection of the null hypothesis $H_0$ according to a calculation of a discriminating
statistic.
 This will suggest that the original data is very unlikely to have
 been generated by a process consistent with the null hypothesis.
\end{itemize} If the value of the statistic calculated on the
original data set is different from the sets of values obtained on
the surrogate data, we have a clear indication for the rejection of
the null. There are various different nulls, some more composite
than others and each null is usually accompanied by its own
procedure of surrogate data generation. For the purposes of this
paper we followed Refs. \citep{Kantz,Schreiber96a,Schreiber00}. We
thus generated phase-randomized amplitude-adjusted surrogates
(termed ``AAFT") to test for the null hypothesis that the return
sequences were monotonic nonlinear transformation of linearly
filtered noise (which is also maintained as the ``most
interesting"). Such surrogates are expected to exhibit the same
spectral and distributional characteristics as in the original
series, however they are purely linear processes. As a
discriminating statistic we chose the BDS test
\citep{Brock87,BHL91,Brock:91,LEBaron}.

We simulated AAFT surrogate data from the original returns
sequences, and produced the compass roses  for various stocks. An
example of an AAFT surrogate set compass rose for the BP stock is
presented in Fig. \ref{fig:fig3}(c). We can clearly see there that
both the randomly shuffled sequence (Fig. \ref{fig:fig3}(b)) and the
AAFT surrogates loose the compass rose structure whereas the
bootstrapped sequence maintains it (Fig.
\ref{fig:fig3}(d)).\footnote{For a discussion on the differences of
bootstrapping and surrogate data analysis refer to \cite{moore99}.}
This was an initial indication that the results of clustering and
discreteness may not be manifestations of linear-random dynamics.

Following the results of the SDA analysis on the phase portraits, we
chose to test for independence under an SDA framework for a subset
of 53 FTSE100 stocks' returns. We used the BDS test as a
discriminating statistic, and generated the AAFT surrogate sets for
each stock, testing the null at 5\%, 2.5\% and 1\% significance
levels. In tables \ref{tab:table1} and \ref{tab:table2}, we present
the results of the SDA. In table \ref{tab:table1} we quote the
results for a BDS test neighborhood size of $0.5$ times the standard
deviation of each returns sequence, for significance levels
$\alpha=$ $5\%$, $2.5\%$ and $1\%$. The results here refute clearly
the null that the sequences are a monotonic nonlinear transformation
of linearly filtered white noise. This is a strong indication of
absence of linear dynamics and randomness and supports the premise
of nonlinear deterministic complexity in the returns. The results of
table \ref{tab:table2} are also supporting this finding. There we
have provided more detail, checking for neighborhood sizes of
$\epsilon_1 = 0.5 \times s_x$, $\epsilon_2 = 1.0 \times s_x$,
$\epsilon_3 = 1.5 \times s_x$ and $\epsilon_4 = 2.0 \times s_x$,
where $s_x$ denotes the standard deviation of each returns sequence.
The level of significance for table \ref{tab:table2} is
$\alpha=5\%$. The results clearly show that  the above null is
strongly refuted.

\bigskip
\begin{center}
[ Insert Table \ref{tab:table1} about here. ]
\end{center}
\bigskip

\bigskip
\begin{center}
[ Insert Table \ref{tab:table2} about here. ]
\end{center}
\bigskip

\bigskip
\begin{center}
[ Insert Table \ref{tab:table3} about here. ]
\end{center}
\bigskip

Since the results of SDA where  pointing towards more complex,
nonlinear dynamics (possibly deterministic) we tested as a next
step, the returns sequences after these have been filtered for noise
reduction. For each stock returns sequence we produced a filtered
version,  using the Waveshrink \citep{donoho95adapting} approach. We
then produced AAFT surrogates and tested for $\alpha=2.5\%$
significance level. In table \ref{tab:table3} we produce the results
for the BP stock, where the Waveshrink
\citep{bru-gao:waveshrink,donoho95adapting} routine has been applied
for a Daubechies 8 (D8) wavelet.\footnote{Choices of different
mother wavelets produced similar results. See also Ref.
\cite{AntoniouVorlow:04}}. Wavelets here are a justified choice in
order to avoid the ``bleaching" of the returns sequences
\citep{Theiler93}, and preserve any delicate deterministic
structures in the DGPs. Our approach is also consistent with Refs.
\cite{capo1,capo2,capo3}. Looking at the values of the BDS statistic
for the original prefiltered sequence and its AAFT surrogates, as
well as the p-value of the statistic for sizes of neighborhood
ranging from $0.5$ to $1.5$ times  the standard deviation, we can
safely reject the null at a $5\%$ significance level. Only for a
size of neighborhood of $2 \times $ standard deviation $\epsilon_4 =
0.0035$ (which is a considerable size), we can reject  the null at a
level of significance of almost $70\%$.

\begin{figure}
  \centering
  \includegraphics*[width=4in]{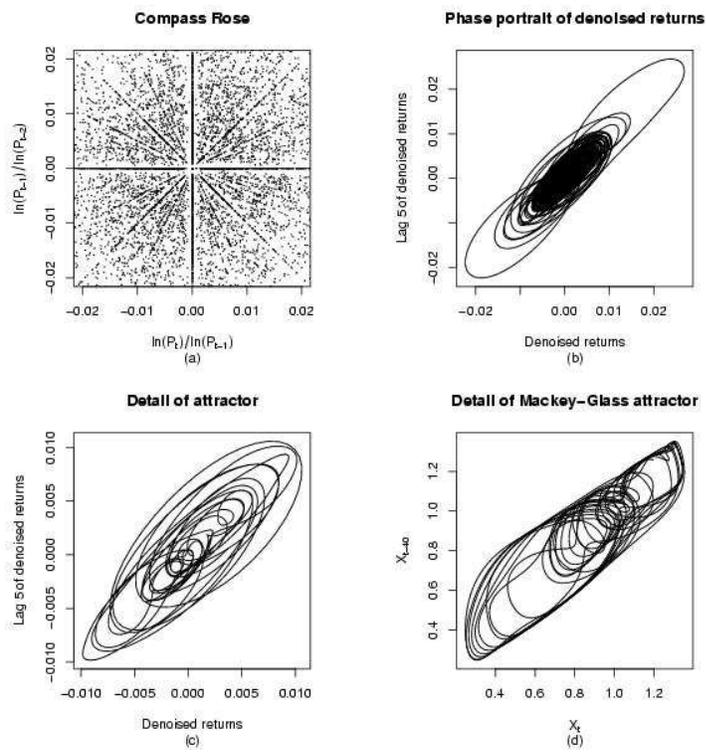}
  \caption{The BP stock returns compass rose (a), details of the denoised BP returns phase portraits (b,c) and the Mackey-Glass attractor (d).}\label{fig:fig1}
\end{figure}

Searching for qualitative evidence of deterministic dynamics and
aperiodic cycles we looked at the phase portraits of the denoised
sequences. For example, in Fig. \ref{fig:fig1} (b) we can clearly
see the phase portrait for the BP denoised returns reveals dynamics
that are similar to chaotic attractors. A detail of the core of the
phase portrait in Fig. \ref{fig:fig1} (c) exhibits dynamics that are
very similar to that of the Mackey-Glass attractor \citep{Mackey77}
in Fig. \ref{fig:fig1} (d). This appears to be in line with
\cite{Kyrtsou1,Kyrtsou2}.

\begin{figure}
  \centering
  \includegraphics*[width=4in]{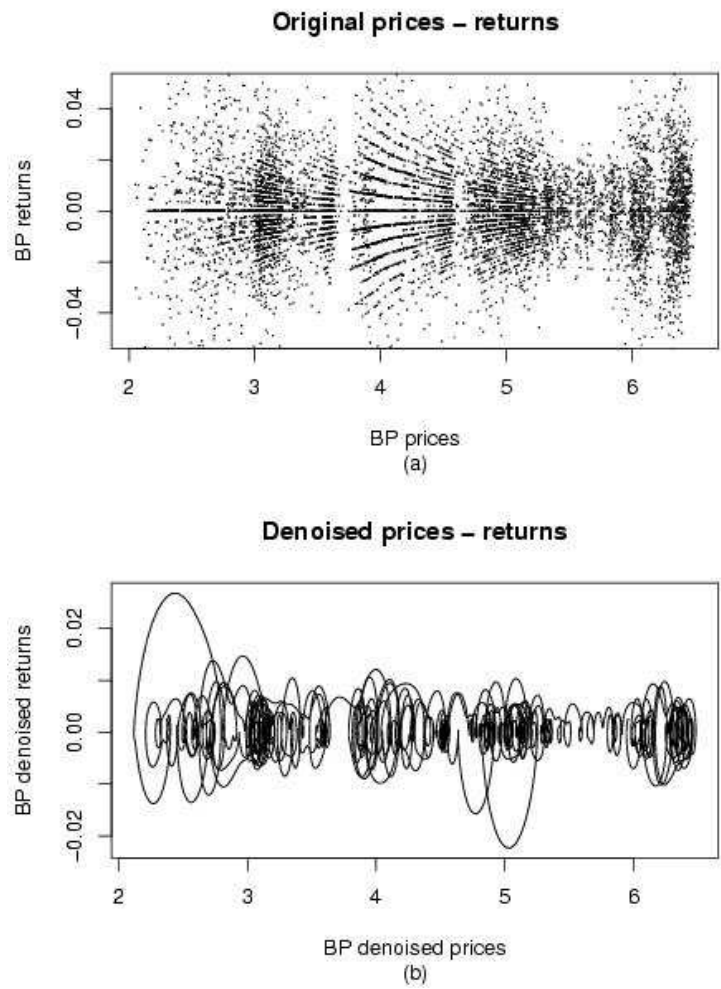}
  \caption{Phase diagrams of the original (a) and denoised (b) BP prices-returns.}\label{fig:fig2}
\end{figure}

\begin{figure}
  \centering
  \includegraphics*[width=4in]{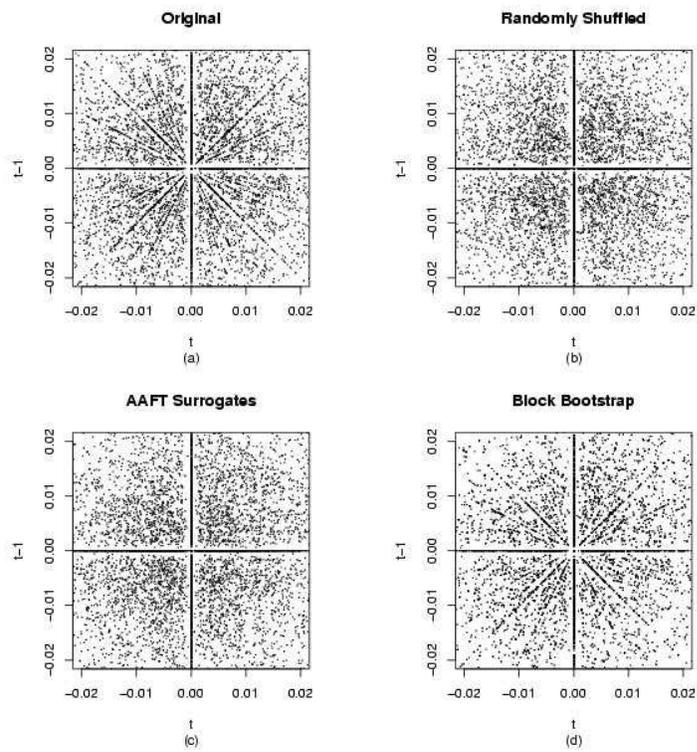}
  \caption{Details of compass roses of the original (a), randomly shuffled (b), AAFT surrogate (c) and bootstrapped (d) BP returns sequences.}\label{fig:fig3}
\end{figure}

\bigskip
\begin{center}
[ Insert figure \ref{fig:fig2} about here. ]
\end{center}
\bigskip

\bigskip
\begin{center}
[ Insert figure \ref{fig:fig3} about here. ]
\end{center}
\bigskip

Another interesting diagram that reveals the effects of stock price
clustering and discreteness is depicted in Fig. \ref{fig:fig2} (a).
There we have plotted the prices of BP stock against the
corresponding logarithmic returns. We can clearly see patterns of
correlation and anticorrelation in the same diagram.  This is the
first time such patterns have been revealed in financial literature
and they need to be investigated further. In nonlinear science, the
phase portraits (i.e., the compass rose) are usually called ``delay
plots" whereas the plot of a sequence of prices from a function
gainst its first derivative are called ``phase plots". Thus the
diagram in Fig. \ref{fig:fig2} (a) could be loosely termed as a
phase plot. If we generate the same kind of display for the denoised
sequences (in this case for the BP stock), we see clearly the
cyclical but aperiodic behavior observed in the phase portraits also
repeated here (Fig. \ref{fig:fig2} (b)).

The results lead us to deduce that the presence of chaotic dynamics
can not be excluded. Such a statement though should also involve the
calculation of certain invariant measures that characterize chaos
(such as entropy or dimension based statistics). Moreover, these
results should also be backed by a suitable SDA testing exercise. We
retain this as a strategy for future research.  It would also be
interesting to observe if these smoother though irregular cyclical
dynamics revealed in this paper are irrespective of the noise
reduction technique (i.e., robust under different noise reduction
techniques).

\section{Conclusions and future research}\label{sec:conclusions}

We have investigated the dynamics of sequences of daily closing
prices and the corresponding returns for stocks traded in the London
Stock Exchange in the last three decades, as these are observed
through the compass rose phase portraits. Our results suggest that
the amount of noise inherent in the examined sequences may be
covering more ``interesting" dynamics. Using wavelet based noise
reduction techniques we filtered the return sequences only to
uncover a strong aperiodic nonlinear behavior, characteristic of
many phenomena that are governed by complex deterministic dynamics.
The SDA hypothesis testing framework employed here also suggests the
absence of stochastic randomness and linear dynamics for both
original and denoised returns sequences. Our results show that the
apparently random dynamics and discreteness observed in closing
price sequences, may conceal via the generation of noise in the
returns, a more delicate structure and aperiodic cyclical dynamics.
However, further research in needed to maintain the hypothesis of
nonlinear determinism in stock price time series dynamics.

\newpage


\clearpage

\pagestyle{empty}

\begin{table}[ht]
\begin{center}
\caption{Surrogate Data Analysis results on actual returns for 53
companies in the FTSE100. Discriminating statistic: BDS test.
Neighbourhood size $\epsilon = 0.5 \times s_x$, where $s_x =$
standard deviation of $x$. Biases and standard errors (s.e.)
reported for significance levels $\alpha = 5\%,$ $2.5\%$ and $1\%$.}
\label{tab:table1}
\begin{tiny}
\begin{tabular}{lrrrrrrr}
\hline & & \multicolumn{2}{c}{$\alpha=5\%$} & \multicolumn{2}{c}{$\alpha=2.5\%$} & \multicolumn{2}{c}{$\alpha=1\%$} \\
\hline & BDS Statistic & bias & s.e. & bias & s.e. & bias & s.e. \\
\hline \hline
FTSE ALL SHARE - PRICE INDEX    &   27.11   &   $-$25.29    &   1.01    &   $-$25.38    &   1.14    &    $-$25.13   &    1.05 \\
FTSE 100 - PRICE INDEX  &   31.9    &   $-$31.61    &   1.24    &   $-$31.41    &   1.05    &    $-$31.82   &    0.91 \\
ALLIED DOMECQ   &   18.68   &   $-$18.64    &   1.08    &   $-$18.83    &   1.1 &    $-$18.65   &    0.94 \\
AMVESCAP    &   32.51   &   $-$32.38    &   0.7 &   $-$31.86    &   0.98    &    $-$32.25   &    0.90 \\
ASSD.BRIT.FOODS &   28.44   &   $-$28.44    &   0.96    &   $-$28.26    &   1.01    &    $-$28.35   &    0.90 \\
AVIVA   &   22.57   &   $-$22.18    &   0.98    &   $-$22.43    &   1.06    &    $-$22.45   &    1.03 \\
BARCLAYS    &   23.1    &   $-$22.29    &   1.06    &   $-$22.40    &   1.02    &    $-$22.45   &    0.98 \\
BOC GROUP   &   23.19   &   $-$23.19    &   1.11    &   $-$23.16    &   1.01    &    $-$22.97   &    1.03 \\
BOOTS GROUP &   19.99   &   $-$19.55    &   1.27    &   $-$19.45    &   1.06    &    $-$19.46   &    1.04 \\
BP  &   17.27   &   $-$16.95    &   1.13    &   $-$16.97    &   0.9 &    $-$17.03   &    1.08 \\
BRIT.AMERICAN TOBACCO   &   17.76   &   $-$17.43    &   1.23    &   $-$17.57    &   1.12    &    $-$17.53   &    0.97 \\
BRITISH LAND    &   36.57   &   $-$36.36    &   1.27    &   $-$36.19    &   0.97    &    $-$36.57   &    1.09 \\
BUNZL   &   25.95   &   $-$24.82    &   0.78    &   $-$24.68    &   1.32    &    $-$24.95   &    1.17 \\
CADBURY SCHWEPPES   &   24.89   &   $-$24.55    &   1.26    &   $-$24.44    &   0.93    &    $-$24.36   &    1.00 \\
DAILY MAIL 'A'  &   34.58   &   $-$34.10    &   1.19    &   $-$34.21    &   1.22    &    $-$34.12   &    1.04 \\
DIAGEO  &   23.41   &   $-$23.33    &   1.11    &   $-$23.23    &   1.11    &    $-$23.31   &    0.99 \\
DIXONS GP.  &   25.12   &   $-$24.65    &   0.98    &   $-$24.54    &   1.25    &    $-$24.72   &    1.09 \\
EMAP    &   33.47   &   $-$32.82    &   0.9 &   $-$33.04    &   1.13    &    $-$33.04   &    1.01 \\
EXEL    &   31.92   &   $-$30.59    &   1.27    &   $-$30.88    &   1   &    $-$30.97   &    1.05 \\
FOREIGN \& COLONIAL &   25.76   &   $-$25.04    &   0.8 &   $-$25.23    &   1.1 &    $-$25.35   &    1.02 \\
GKN &   22.41   &   $-$22.37    &   0.89    &   $-$22.43    &   1.06    &    $-$22.22   &    1.03 \\
GLAXOSMITHKLINE &   18.48   &   $-$18.05    &   0.94    &   $-$18.04    &   1.14    &    $-$18.27   &    1.13 \\
GRANADA &   31.42   &   $-$30.44    &   1.16    &   $-$30.35    &   0.95    &    $-$30.42   &    1.11 \\
GUS &   43.46   &   $-$43.23    &   1.29    &   $-$43.28    &   0.9 &    $-$43.31   &    1.16 \\
HANSON  &   23.26   &   $-$22.95    &   1.19    &   $-$22.58    &   1   &    $-$22.76   &    0.99 \\
HILTON GROUP    &   22.41   &   $-$22.23    &   1.1 &   $-$22.15    &   1.05    &    $-$22.12   &    0.92 \\
IMP.CHM.INDS.   &   21.56   &   $-$21.43    &   0.96    &   $-$21.50    &   1.04    &    $-$21.35   &    1.05 \\
JOHNSON MATTHEY &   28.6    &   $-$27.32    &   0.9 &   $-$27.76    &   1   &    $-$27.61   &    1.13 \\
LAND SECURITIES &   26.82   &   $-$26.36    &   0.79    &   $-$26.31    &   1.15    &    $-$26.33   &    1.05 \\
LEGAL \& GENERAL    &   24.83   &   $-$24.67    &   1.23    &   $-$24.67    &   1.08    &    $-$24.62   &    1.03 \\
MARKS \& SPENCER GROUP  &   22.22   &   $-$22.10    &   1.22    &   $-$22.25    &   0.9 &    $-$22.26   &    1.06 \\
MORRISON (WM) SPMKTS.   &   22.36   &   $-$21.71    &   1.22    &   $-$21.91    &   1.11    &    $-$21.86   &    1.08 \\
NEXT    &   23.44   &   $-$23.05    &   1.05    &   $-$23.05    &   0.9 &    $-$22.85   &    1.06 \\
PEARSON &   29.34   &   $-$28.47    &   1.05    &   $-$28.56    &   0.78    &    $-$28.63   &    1.15 \\
PROVIDENT FINL. &   32.23   &   $-$31.93    &   1.13    &   $-$31.53    &   1.03    &    $-$31.45   &    1.04 \\
PRUDENTIAL  &   23.3    &   $-$23.13    &   0.87    &   $-$23.12    &   0.95    &    $-$23.14   &    0.96 \\
RECKITT BENCKISER   &   23.43   &   $-$22.62    &   1.29    &   $-$22.62    &   0.76    &    $-$22.45   &    1.00 \\
REED ELSEVIER   &   24.67   &   $-$24.40    &   1.37    &   $-$24.29    &   0.93    &    $-$24.43   &    0.92 \\
RENTOKIL INITIAL    &   29.41   &   $-$28.93    &   0.78    &   $-$29.09    &   1.07    &    $-$28.91   &    1.04 \\
REXAM   &   26.26   &   $-$26.08    &   1.03    &   $-$25.63    &   1.1 &    $-$25.97   &    1.05 \\
RIO TINTO   &   23.14   &   $-$22.48    &   0.8 &   $-$22.58    &   1.21    &    $-$22.52   &    1.15 \\
ROYAL BANK OF SCOTLAND  &   27.18   &   $-$26.85    &   1.11    &   $-$26.83    &   0.92    &    $-$26.81   &    1.11 \\
SAINSBURY (J)   &   27.13   &   $-$26.78    &   0.92    &   $-$26.66    &   1.07    &    $-$26.62   &    0.99 \\
SCHRODERS   &   32.63   &   $-$32.39    &   1.17    &   $-$32.45    &   1.05    &    $-$32.46   &    0.95 \\
SCOT. \& NEWCASTLE  &   28.46   &   $-$27.93    &   1.25    &   $-$28.43    &   1.08    &    $-$28.27   &    1.05 \\
SHELL TRANSPORT \& TRDG. &   24.07   &   $-$23.36    &   0.96    &   $-$23.69    &   0.86    &    $-$23.47   &    1.05 \\
SMITH \& NEPHEW &   28.28   &   $-$28.07    &   0.74    &   $-$27.76    &   0.98    &    $-$27.78   &    0.91 \\
SMITHS GROUP    &   25.67   &   $-$24.30    &   0.99    &   $-$24.00    &   1.14    &    $-$24.07   &    0.89 \\
STD.CHARTERED   &   33.79   &   $-$32.78    &   0.71    &   $-$33.03    &   0.92    &    $-$32.88   &    1.09 \\
TESCO   &   20.95   &   $-$20.58    &   1.12    &   $-$20.53    &   0.78    &    $-$20.49   &    0.92 \\
TOMKINS &   27.42   &   $-$27.36    &   0.95    &   $-$27.57    &   0.89    &    $-$27.42   &    1.05 \\
UNILEVER (UK)   &   23.95   &   $-$23.70    &   0.81    &   $-$23.25    &   1.14    &    $-$23.48   &    1.03 \\
WHITBREAD   &   22.32   &   $-$21.59    &   1.14    &   $-$21.64    &   0.82    &    $-$21.81   &    1.06 \\
WOLSELEY    &   26.37   &   $-$25.03    &   1.2 &   $-$25.16    &   1.13    &    $-$25.50   &    1.00 \\
WPP GROUP   &   34.11   &   $-$33.73    &   0.86    &   $-$33.80    &   0.94    &    $-$33.64   &    0.91 \\
\hline \hline
\end{tabular}
\end{tiny}
\end{center}
\end{table}

\newpage

\begin{table}[ht]
\begin{center}
\caption{Surrogate Data Analysis results on actual returns for 53
companies in the FTSE100. Discriminating statistic: BDS test
(embedding dimension 3). Neighbourhood size $\epsilon_1 = 0.5 \times
s_x$, $\epsilon_2 = 1.0 \times s_x$, $\epsilon_3 = 1.5 \times s_x$
and $\epsilon_4 = 2.0 \times s_x$, where $s_x =$ standard deviation
of $x$. Biases and standard errors reported for significance level
$\alpha = 1\%$ .}\label{tab:table2}
\begin{tiny}
\begin{tabular}{lrrrrrrrrrrrr}
\hline &  \multicolumn{4}{c}{Statistic (BDS)} &
\multicolumn{4}{c}{Bias} & \multicolumn{4}{c}{Standard Error} \\
\hline
Neighbourhood size  & $\epsilon_1$ & $\epsilon_2$ & $\epsilon_3$ & $\epsilon_4$  & $\epsilon_1$ & $\epsilon_2$ & $\epsilon_3$ & $\epsilon_4$  & $\epsilon_1$ & $\epsilon_2$ & $\epsilon_3$ & $\epsilon_4$ \\
\hline \hline
FTSE ALL SHARE - PRICE INDEX    &   20.8    &   23.87   &   26.81   &   28.72   &   $-$18.61    &   $-$21.45    &   $-$24.32    &   $-$26.37    &   1.11    &   1.13    &   1.17    &   1.23    \\
FTSE 100 - PRICE INDEX  &   21.25   &   15.31   &   15.27   &   13.02   &   $-$21.41    &   $-$15.44    &   $-$15.41    &   $-$13.09    &   1.1 &   0.92    &   0.87    &   0.99    \\
ALLIED DOMECQ   &   13.73   &   15.09   &   16.38   &   17.17   &   $-$13.72    &   $-$15.01    &   $-$16.25    &   $-$16.99    &   1.02    &   0.97    &   0.99    &   1   \\
AMVESCAP    &   25.6    &   24.1    &   22.33   &   21.13   &   $-$25.13    &   $-$23.55    &   $-$21.74    &   $-$20.52    &   0.95    &   0.89    &   0.91    &   1   \\
ASSD.BRIT.FOODS &   22.73   &   24.5    &   23.7    &   22.53   &   $-$22.50    &   $-$24.33    &   $-$23.49    &   $-$22.41    &   1.06    &   1   &   1.04    &   1.08    \\
AVIVA   &   18.31   &   19.37   &   19.79   &   20.26   &   $-$18.44    &   $-$19.37    &   $-$19.75    &   $-$20.18    &   1.11    &   1.18    &   1.14    &   1.11    \\
BARCLAYS    &   18.25   &   20.46   &   22.67   &   23.97   &   $-$17.82    &   $-$19.74    &   $-$21.82    &   $-$23.12    &   1.04    &   1.09    &   1.03    &   0.98    \\
BOC GROUP   &   18.3    &   17.06   &   16.67   &   16.13   &   $-$18.25    &   $-$16.93    &   $-$16.48    &   $-$16.00    &   0.91    &   1.02    &   1.06    &   1.01    \\
BOOTS GROUP &   15.77   &   16  &   17.12   &   18.31   &   $-$15.19    &   $-$15.32    &   $-$16.44    &   $-$17.74    &   1   &   1.01    &   0.99    &   0.99    \\
BP  &   12.89   &   13.62   &   13.8    &   14.07   &   $-$12.68    &   $-$13.36    &   $-$13.44    &   $-$13.69    &   1.02    &   0.94    &   0.93    &   0.95    \\
BRIT.AMERICAN TOBACCO   &   14.28   &   16.3    &   16.96   &   17.67   &   $-$14.50    &   $-$16.34    &   $-$16.91    &   $-$17.55    &   1.07    &   1.21    &   1.19    &   1.13    \\
BRITISH LAND    &   28.42   &   29.97   &   30.8    &   32.2    &   $-$28.28    &   $-$29.80    &   $-$30.44    &   $-$31.82    &   1.01    &   1.01    &   1.01    &   1.05    \\
BUNZL   &   21.76   &   19.84   &   18.87   &   15.15   &   $-$20.55    &   $-$18.16    &   $-$17.00    &   $-$13.30    &   1.03    &   0.9 &   0.99    &   1.19    \\
CADBURY SCHWEPPES   &   20.53   &   22.39   &   23.32   &   23.81   &   $-$20.00    &   $-$21.82    &   $-$22.68    &   $-$23.11    &   0.97    &   1.05    &   1.04    &   1.1 \\
DAILY MAIL `A'  &   28.24   &   28.42   &   26.45   &   22.3    &   $-$27.93    &   $-$27.91    &   $-$25.90    &   $-$21.72    &   1.14    &   1.26    &   1.22    &   1.07    \\
DIAGEO  &   18.05   &   18.36   &   18.96   &   19.04   &   $-$17.58    &   $-$17.75    &   $-$18.32    &   $-$18.45    &   1.18    &   1.15    &   1.17    &   1.12    \\
DIXONS GP.  &   20.6    &   20.62   &   20.17   &   19.48   &   $-$19.95    &   $-$19.95    &   $-$19.59    &   $-$19.08    &   1.14    &   1.09    &   0.95    &   0.84    \\
EMAP    &   27.63   &   22.23   &   19.17   &   17.41   &   $-$27.25    &   $-$21.93    &   $-$18.94    &   $-$17.17    &   1.04    &   1.03    &   1.08    &   1.08    \\
EXEL    &   25.56   &   23.08   &   19.81   &   17.98   &   $-$24.13    &   $-$21.34    &   $-$18.02    &   $-$16.34    &   0.78    &   1.03    &   1.18    &   1.2 \\
FOREIGN \& COLONIAL &   20.47   &   19.83   &   21.56   &   22.04   &   $-$19.97    &   $-$19.22    &   $-$20.87    &   $-$21.46    &   0.96    &   0.97    &   0.98    &   0.99    \\
GKN &   17.02   &   18.68   &   18.42   &   17.65   &   $-$16.55    &   $-$18.15    &   $-$17.85    &   $-$17.07    &   1.04    &   1.04    &   1.13    &   1.16    \\
GLAXOSMITHKLINE &   14.12   &   15.45   &   16.11   &   16.3    &   $-$13.89    &   $-$15.17    &   $-$15.77    &   $-$15.95    &   0.89    &   0.89    &   0.82    &   0.93    \\
GRANADA &   24.34   &   26.02   &   24.88   &   22.38   &   $-$23.31    &   $-$24.73    &   $-$23.46    &   $-$21.01    &   1.05    &   1.06    &   1.06    &   0.99    \\
GUS &   29.9    &   25.99   &   24.7    &   22.03   &   $-$29.71    &   $-$25.71    &   $-$24.45    &   $-$21.82    &   1.02    &   1.03    &   0.92    &   0.87    \\
HANSON  &   19.56   &   19.93   &   19.53   &   18.07   &   $-$18.57    &   $-$18.89    &   $-$18.55    &   $-$17.22    &   1.07    &   1.13    &   1.1 &   1.11    \\
HILTON GROUP    &   17.77   &   18.17   &   18.48   &   19.22   &   $-$17.17    &   $-$17.28    &   $-$17.44    &   $-$18.27    &   1.02    &   1.01    &   1.04    &   1.03    \\
IMP.CHM.INDS.   &   17.11   &   18.92   &   19.68   &   19.58   &   $-$17.28    &   $-$18.84    &   $-$19.45    &   $-$19.38    &   1.05    &   1.01    &   1.03    &   1.05    \\
JOHNSON MATTHEY &   22.63   &   21.34   &   20.24   &   17.27   &   $-$21.60    &   $-$19.93    &   $-$18.77    &   $-$15.94    &   1.09    &   1.13    &   1.08    &   0.99    \\
LAND SECURITIES &   21.54   &   22.95   &   24.18   &   25.44   &   $-$20.93    &   $-$22.33    &   $-$23.57    &   $-$24.93    &   0.9 &   0.9 &   0.87    &   0.92    \\
LEGAL \& GENERAL    &   19.17   &   20.94   &   22.65   &   23.96   &   $-$18.98    &   $-$20.72    &   $-$22.36    &   $-$23.57    &   1.13    &   1.08    &   1.04    &   1.07    \\
MARKS \& SPENCER GROUP  &   17.69   &   19.03   &   20.36   &   20.98   &   $-$17.93    &   $-$19.25    &   $-$20.58    &   $-$21.24    &   0.95    &   0.92    &   0.85    &   0.8 \\
MORRISON (WM) SPMKTS.   &   19.14   &   18.52   &   17.87   &   15.61   &   $-$18.71    &   $-$17.76    &   $-$17.14    &   $-$15.00    &   1.16    &   0.93    &   0.94    &   0.97    \\
NEXT    &   17.35   &   18.96   &   21.43   &   21.21   &   $-$16.96    &   $-$18.41    &   $-$20.83    &   $-$20.68    &   1   &   1.06    &   1.11    &   1.08    \\
PEARSON &   23.95   &   24.97   &   23.55   &   22.58   &   $-$22.81    &   $-$23.65    &   $-$22.22    &   $-$21.38    &   0.92    &   0.91    &   0.94    &   1.04    \\
PROVIDENT FINL. &   23.8    &   22.06   &   20.06   &   18.6    &   $-$23.10    &   $-$21.24    &   $-$19.10    &   $-$17.76    &   1.19    &   1.08    &   0.96    &   0.94    \\
PRUDENTIAL  &   18.21   &   19.09   &   21.01   &   22.67   &   $-$17.52    &   $-$18.38    &   $-$20.40    &   $-$22.19    &   0.86    &   0.92    &   1.02    &   1.1 \\
RECKITT BENCKISER   &   19.09   &   20.73   &   21.89   &   22.29   &   $-$18.03    &   $-$19.43    &   $-$20.45    &   $-$20.91    &   1.06    &   0.93    &   0.92    &   0.97    \\
REED ELSEVIER   &   19.57   &   19.57   &   19.44   &   18.94   &   $-$19.15    &   $-$19.02    &   $-$18.87    &   $-$18.47    &   0.79    &   0.9 &   1.04    &   1.03    \\
RENTOKIL INITIAL    &   23.14   &   21.06   &   21.56   &   20.11   &   $-$22.77    &   $-$20.42    &   $-$20.90    &   $-$19.50    &   0.84    &   0.99    &   1.06    &   1.02    \\
REXAM   &   20.18   &   19.56   &   18.95   &   18.24   &   $-$19.89    &   $-$19.18    &   $-$18.50    &   $-$17.66    &   0.87    &   0.95    &   1.02    &   1.02    \\
RIO TINTO   &   17.96   &   18.65   &   18.84   &   18.71   &   $-$17.02    &   $-$17.46    &   $-$17.57    &   $-$17.42    &   0.85    &   0.84    &   0.96    &   1.06    \\
ROYAL BANK OF SCOTLAND  &   21.44   &   21.5    &   22.06   &   21.63   &   $-$21.15    &   $-$21.16    &   $-$21.64    &   $-$21.08    &   1.06    &   1.06    &   1.02    &   0.99    \\
SAINSBURY (J)   &   21.44   &   22.58   &   23.46   &   22.49   &   $-$21.16    &   $-$22.18    &   $-$22.85    &   $-$21.79    &   0.9 &   0.91    &   0.96    &   1.04    \\
SCHRODERS   &   25.54   &   27.25   &   26.83   &   25.09   &   $-$25.43    &   $-$27.05    &   $-$26.39    &   $-$24.92    &   1.17    &   1   &   1.08    &   0.91    \\
SCOT. \& NEWCASTLE  &   21.71   &   21.8    &   22.32   &   22.31   &   $-$21.38    &   $-$21.37    &   $-$21.82    &   $-$21.76    &   0.86    &   0.86    &   0.98    &   1.06    \\
SHELL TRANSPORT \& TRDG.    &   19.4    &   20.45   &   20.78   &   20.38   &   $-$19.14    &   $-$20.11    &   $-$20.39    &   $-$20.00    &   1   &   1   &   1.17    &   1.24    \\
SMITH \& NEPHEW &   22.05   &   22.23   &   20.99   &   20.91   &   $-$21.76    &   $-$21.77    &   $-$20.43    &   $-$20.34    &   1.12    &   1.11    &   1.03    &   1.06    \\
SMITHS GROUP    &   20.72   &   20.23   &   19.2    &   18.16   &   $-$18.91    &   $-$17.98    &   $-$16.78    &   $-$15.94    &   1.02    &   1.11    &   1.09    &   1.03    \\
STD.CHARTERED   &   26.61   &   25.67   &   24.1    &   21.95   &   $-$25.44    &   $-$24.23    &   $-$22.63    &   $-$20.53    &   0.97    &   1.14    &   1.17    &   1.21    \\
TESCO   &   16.39   &   15.57   &   15.6    &   16.05   &   $-$16.14    &   $-$15.20    &   $-$15.06    &   $-$15.44    &   1   &   1.18    &   1.19    &   1.22    \\
TOMKINS &   21.92   &   14.76   &   15.41   &   13.84   &   $-$21.79    &   $-$14.55    &   $-$15.25    &   $-$13.67    &   1.07    &   0.92    &   1.02    &   0.95    \\
UNILEVER (UK)   &   19.48   &   20.62   &   20.74   &   19.9    &   $-$18.71    &   $-$19.76    &   $-$19.96    &   $-$19.30    &   0.83    &   0.88    &   0.94    &   0.92    \\
WHITBREAD   &   16.91   &   17.65   &   17.6    &   17.21   &   $-$16.26    &   $-$16.79    &   $-$16.73    &   $-$16.37    &   0.89    &   0.84    &   0.9 &   0.94    \\
WOLSELEY    &   20.51   &   19.06   &   17.85   &   16.49   &   $-$19.71    &   $-$17.82    &   $-$16.45    &   $-$15.20    &   1.14    &   1.01    &   0.88    &   0.88    \\
WPP GROUP   &   27.81   &   22.97   &   20.73   &   23.13   &   $-$27.38    &   $-$22.23    &   $-$19.98    &   $-$22.17    &   1.06    &   1.02    &   1.3 &   1.34    \\
\hline \hline
\end{tabular}
\end{tiny}
\end{center}
\end{table}

\newpage

\begin{table}[ht]
\begin{center}
\caption{Surrogate Data Analysis results on D8 pre-filtered BP
returns. Discriminating statistic: BDS test (embedding dimension 2).
Neighbourhood size $\epsilon_1 = 0.5 \times s_x$, $\epsilon_2 = 1.0
\times s_x$, $\epsilon_3 = 1.5 \times s_x$ and $\epsilon_4 = 2.0
\times s_x$, where $s_x =$ standard deviation of
$x$.}\label{tab:table3}
\begin{tiny}
\begin{tabular}{lrrrr}
\hline & \multicolumn{4}{c}{Neighbourhood Size}  \\ \hline
  set & $\epsilon_1=9e-04$ & $\epsilon_1=0.0017$ & $\epsilon_2=0.0026$ & $\epsilon_4=0.0035$ \\
\hline  \hline
1 & 922.89 & 444.06 & 316.55 & 280.90 \\
2 & 837.02 & 417.54 & 300.76 & 266.65 \\
3 & 934.51 & 446.17 & 317.84 & 281.37 \\
4 & 933.59 & 446.77 & 318.31 & 281.97 \\
5 & 880.90 & 430.70 & 308.58 & 273.29 \\
6 & 936.90 & 446.80 & 318.40 & 281.44 \\
7 & 889.16 & 432.69 & 309.08 & 272.54 \\
8 & 928.31 & 444.19 & 316.63 & 279.94 \\
9 & 916.11 & 441.64 & 316.03 & 280.41 \\
10 & 881.64 & 431.72 & 308.36 & 273.35 \\
11 & 916.71 & 441.76 & 315.27 & 279.04 \\
12 & 932.64 & 446.76 & 318.92 & 282.99 \\
13 & 832.50 & 417.22 & 300.04 & 266.62 \\
14 & 932.27 & 446.08 & 318.41 & 282.52 \\
15 & 925.76 & 444.06 & 317.08 & 280.87 \\
16 & 941.02 & 447.37 & 318.04 & 279.56 \\
17 & 913.54 & 440.37 & 314.19 & 278.49 \\
18 & 888.79 & 433.23 & 309.91 & 274.35 \\
19 & 935.24 & 446.81 & 318.27 & 281.28 \\
20 & 831.79 & 416.32 & 299.63 & 265.58 \\
21 & 799.68 & 406.98 & 292.95 & 259.33 \\
22 & 832.68 & 416.32 & 298.47 & 264.23 \\
23 & 865.91 & 425.52 & 305.02 & 269.74 \\
24 & 884.16 & 432.43 & 310.04 & 275.97 \\
25 & 882.36 & 431.60 & 309.22 & 274.49 \\
26 & 872.69 & 427.79 & 306.57 & 271.06 \\
27 & 903.62 & 437.21 & 312.30 & 276.44 \\
28 & 943.29 & 449.77 & 320.33 & 284.10 \\
29 & 927.47 & 444.60 & 317.18 & 280.81 \\
30 & 897.28 & 435.32 & 310.77 & 274.88 \\
31 & 931.34 & 446.29 & 318.95 & 283.01 \\
32 & 892.41 & 434.36 & 311.11 & 276.10 \\
33 & 921.26 & 441.78 & 315.33 & 277.76 \\
34 & 882.30 & 432.09 & 309.36 & 274.94 \\
35 & 938.17 & 448.33 & 319.22 & 282.41 \\
36 & 935.55 & 447.50 & 318.72 & 282.32 \\
37 & 809.99 & 409.72 & 295.84 & 262.73 \\
38 & 919.59 & 442.14 & 315.87 & 279.80 \\
39 & 885.79 & 432.02 & 309.67 & 274.05 \\
40 & 877.49 & 430.08 & 308.29 & 273.56 \\
\hline

\hline Original &  1793.68 &  567.17 &  337.70  & 273.50
\\
\hline Significance &  23.07 &  11.43 & 3.61 & 0.40 \\
\hline p-value & 0.00 &  0.00 &  0.00 &  0.69
 \\ \hline \hline
\end{tabular}
\end{tiny}
\end{center}
\end{table}

\clearpage \newpage

\newpage

\newpage

\end{document}